# Canting angle behavior of magnetic moments in Y- substituted $Tb_2BaNiO_5$ and its relevance for magnetoelectric coupling

Ram Kumar[1,2], S. Rayaprol[3], A. Hoser,[4] and E. V. Sampathkumaran[1,5]

[1]*Tata Institute of Fundamental Research, Homi Bhabha Road, Colaba, Mumbai 400005, India*
[2]*Maryland Quantum Materials Center and Department of Physics, University of Maryland, College Park, Maryland 20742, USA*
[3]*UGC-DAE Consortium for Scientific Research, Mumbai Centre, BARC Campus, Trombay, Mumbai 400085, India*
[4]*Helmholtz-Zentrum Berlin für Materialien und Energie, Berlin, D-14109, Germany*
[5]*Homi Bhabha Centre for Science Education, Tata Institute of Fundamental Research, V. N. Purav Marg, Mankhurd, Mumbai, 400088 India*

The Haldane-spin chain compound, $Tb_2BaNiO_5$ has been known to be an exotic multiferroic system, exhibiting antiferromagnetic anomalies at $T_{N1}$= 63 K and $T_{N2}$= 25 K, with ferroelectricity appearing below $T_{N2}$ only. Previous reports in addition established that, interestingly, Tb ions play a direct and decisive role to lead to multiferroic properties with a critical canting angle of magnetic moments, unlike other well-known multiferroics. Here, we report the results of temperature dependent neutron powder diffraction studies on $Tb_{2-x}Y_xBaNiO_5$, to get an insight into the critical canting angle for multiferroic behavior. While multiferroic transition temperature decreases linearly with Y concentration, there is an abrupt drop of relative canting angle (of Tb and Ni magnetic moments) with respect to that in parent compound for an initial substitution of x= 0.5 in the multiferroic region, without any notable change thereafter. We therefore infer that this critical canting angle is made up of two components - cooperative (long-range) and local (short-range) contributions.

**E-mail addresses of authors:** ramasharamyadav@gmail.com, rayaprol@gmail.com, hoser@helmholtz-berlin.de**,** sampathev@gmail.com

**Introduction**

In recent times, there is a considerable interest to identify the materials exhibiting spin-driven magnetoelectric coupling and for finding mechanisms and routes to engineer such materials [1–19] considering application potential. In this respect, the Haldane spin-chain family of the type, $R_2BaNiO_5$ ($R$ = rare-earth) [20, 21] is of great interest, as a variety of magnetodielectric (MDE) and multiferroic behaviors have been found among these compounds. These compounds form in an orthorhombic structure (space group *Immm*) [22-26], with Ni-O spin chains running along the *a*-axis. Among these, the Tb compound has been found to be an extraordinary system [23-27]: Two antiferromagnetic (AFM) features, one at $T_{N1}$ = 63 K and the other at $T_{N2}$ = 25 K have been established for this compound, with the Néel temperature ($T_{N1}$) being the highest within this family due to anisotropic 4*f* orbital; MDE coupling is also the highest (about 18%) among polycrystalline materials not only within this family, but also with respect to other materials at the time of its publication [23]. An intriguing finding brought out by neutron powder diffraction (NPD) studies is that the mutually canted Tb and Ni spins (which are collinear within their sublattices) below $T_{N1}$ gives rise to the electric polarization (*P*) below $T_{N2}$ only, when the relative canting angle ($\Delta\theta$), defined as the difference of the canting angles of Tb and Ni moments with respect to *c*-axis, exceeds a critical value [24]. By this conclusion, we do not claim that these angles are order parameters of the multiferroic transition, but noted a correlation. The disappearance [25] of spontaneous electric polarization for a small replacement of Ba by Sr (10%), which does not involve disturbing the periodicity of Tb sublattice, and a drastic fall [27] in the canting angle below that for the parent compound further supported the idea of critical canting angle for multiferroicity. Such a condition of critical canting angle cannot be accountable within hitherto known theories including conventional Dzyaloshinski-Moriya interaction (DMI) based models [14].

We consider it important to go deeper into the origin of critical canting angle to enable further knowledge to engineer materials for applications. This prompted us to perform NPD studies on a family involving disruption of Tb sublattice (in contrast to Sr-doping in which Tb sublattice is intact), viz., Y-substituted series, $Tb_{2-x}Y_xBaNiO_5$. The bulk measurements on this series established [26] that both $T_{N1}$ and $T_{N2}$ fall linearly with increasing Y-doping, with $T_{N2}$ tracking multiferroic temperature till the lowest measured Tb-composition of x= 1.5. This finding clearly revealed the role of Tb 4*f* orbital on the onset of multiferroic anomaly, unlike in most other well-known multiferroics in which transition metal ions are responsible for multiferroicity. The results reported here suggest that both cooperative and local effects contribute to decide critical canting angle.

**Experimental details**

Polycrystalline single-phase samples of $Tb_{2-x}Y_xBaNiO_5$ (x = 0.5, 1.0 and 1.5) were prepared by a standard solid-state reaction method as described earlier [26], starting with stoichiometric amounts of $Tb_2(CO_3)_2 \cdot nH_2O$ (99.9%), $Y_2O_3$ (99.9%), NiO (99.995%), and $BaCO_3$ (99.997%). The specimens were characterized by x-ray diffraction (XRD) and also by the temperature (*T*) and field (*H*) dependent *dc* magnetization measurements to make sure that the results of Ref. 26 are reproduced on this batch of samples [26]. NPD patterns were obtained using the wavelength, λ=2.451 Å, at several temperatures on E6 diffractometer at Helmholtz-Zentrum Berlin (HZB). A few grams of powder were packed in a vanadium can and attached to a variable temperature insert, which was then loaded into a standard liquid helium orange cryostat for obtaining *T*-dependent neutron diffraction patterns. The data for each spectrum was collected in 48 scan steps of the 2 two-dimensional area detectors. The total angular range covered (in 2θ) is from 4° to 136° [28]. The specimen was initially cooled to 1.8 K and the diffraction patterns were recorded at selected temperatures during warming cycle. The NPD data was refined for crystallographic as well as magnetic structures by Rietveld refinement method using the FullProf suite programs [29].

**Results and discussion**

We now present how the magnetic structure and, more importantly the canting angles of the magnetic moments of Tb 4*f* and Ni 3*d* with respect to *c*-axis, and their difference |Δθ|, differ in these Y substituted $Tb_2BaNiO_5$ from the parent Tb compound. Using the Rietveld refinement model employed to

XRD patterns in the paramagnetic state (to the *Immm* space group), we have refined the magnetic structures below magnetic ordering. The model is the same as that used for $Tb_2BaNiO_5$ [24]. Figs. 1a-c, depicts the observed ($I_{obs}$) and calculated profile ($I_{cal}$) of NPD patterns for x= 0.5 and 1.0 at *T* = 70, 40, and 2 K, and for x= 1.5, at *T* = 40, 20, and 2 K. In the magnetically ordered state (that is, below $T_{N1}$ ~ 54 K, 43 K and 29 K respectively for x= 0.5, 1.0 and 1.5), e.g., for *T* = 40 K for x= 0.5 and 1.0 and *T*= 20 K for x=1.5, there are a few new Bragg peaks while comparing with the pattern in the paramagnetic state (say, at 70 K). There is also a gradual increase in the intensity of magnetic Bragg peaks (with respect to nuclear Bragg peaks), as Tb and Ni moments gradually align with the lowering the temperature. The magnetic peaks are marked by asterisks in these figures. With a further lowering of temperature below respective $T_{N2}$ (i.e., 21 K, 14 K and 10 K), there is no indication for any change in the magnetic structure. Such NPD patterns can be fitted to the propagation vector, $k = (½, 0, ½)$ at all temperatures in the magnetically ordered state, also considering the crystallographic (nuclear) Bragg peaks. The magnetic structures obtained by fitting are typical of that shown in Fig. 1d for x =1.5 composition at 2 K; one can visualize two interpenetrating antiferromagnetic sublattices (of Tb and Ni ions) canted with each other, as in the case of the parent compound. The Tb moment lies almost along *c*-axis (i.e., canting angle "$\theta$" is very small) around $T_{N1}$, increasing with decreasing temperature, whereas the canting angle for Ni moment undergoes a relatively very weak decrease with *T* (Fig. 2). In sharp contrast to this, in the parent compound, it is known that the canting angle of Ni moment undergoes a more pronounced change with respect to that of Tb [24]. The magnetic moments gradually increase below respective $T_{N2}$, tending to saturate at low temperatures in the close vicinity of $T_{N2}$ (Fig. 3a). The plots of the x and z components (along *a* and *c* axis respectively) of Tb and Ni magnetic moments as a function of temperature are shown in Fig. 3b Interestingly, there is a noticeable difference in the values of the x-component (Tb_x) and z-component (Tb_z) of the Tb magnetic moment in the *T* range $T_{N2}<T<T_{N1}$, whereas x and z components of Ni overlap in the same *T*-range. This is consistent with the observation that the temperature dependence of the canting angle of Tb moments is larger than that of Ni. The most important finding is that the relative canting angle ($\Delta\theta$) of the magnetic moments of Tb and Ni throughout the temperature range is below the critical $\Delta\theta$, known for the parent compound (around 15º), despite that all these Y substituted specimens are also multiferroic. Clearly, therefore, the critical $\Delta\theta$ required to trigger ferroelectricity gets altered when the Tb sublattice is disturbed. However, the plots appear so smeared due to magnetic inhomogeneity induced by the random occupation of (non-magnetic) Y at the (magnetic) Tb site that the jump is not seen. A comparison of the behavior of dielectric constant and electric polarization (*P*) [26] on the one hand and the $\Delta\theta$ on the other is shown in Fig. 4. For the benefit of the reader, the derivatives of the magnetic susceptibility reported in Ref. 26 are also shown, and it is well-known that, in this family of Haldane spin-chain compounds, a change of slope only would be observed at the magnetic transitions; the transitions are much more marked in heat-capacity data. This figure suggests that the values of $\Delta\theta$ above which ferroelectricity arises falls to a value of about 7 to 8º for x= 0.5, and remains essentially the same for other compositions as well. *This is an intriguing finding, as $\Delta\theta$ is not falling gradually with x.* Such an observation implies that a small disruption of Tb sublattice dramatically reduces the value $\Delta\theta$, indicating the role of cooperative magnetic contribution to determine the value of critical angle. The fact that, even for x= 1.5, the value is the same as that for x= 0.5, implies that local effects also play a role. Needless to emphasize that this substitutional study (Y for Tb) does not involve any electron/hole doping with both Y and Tb being in the tri-positive charge state and so the conclusions are free from any possible change in the oxygen content. This is apart from the fact that all the compositions were synthesized under identical heat-treatment conditions. The lattice strain is also negligible, as the ionic radius of $Y^{3+}$ is marginally smaller than that of $Tb^{3+}$, which is evident from the fact that the lattice constants variation with x is also very weak (only in the second decimal), and the linear variation of the lattice constants with x without any non-monotonous tendency also establishes constancy of oxygen content across the series. This is further confirmed by very good fitting of the neutron diffraction patterns in the paramagnetic state for the oxygen stoichiometry of the compositions.

Finally, we may add that there are some solid-state phenomena in which one can see the interplay between local and cooperative effects. For instance, in the field of the Kondo lattices, for example, among Ce-based intermetallics, following logarithmic increase of electrical resistivity with decreasing temperature, due to on-site Kondo effect there is a downturn due to coherent scattering of the conduction electrons by the periodically placed Kondo centres; a dilution of the Ce sublattice by a non-magnetic rare-earth gradually suppresses the temperature at which this downturn due to the Kondo-coherence occurs; this downturn finally vanishes beyond certain concentration of the non-magnetic dopant, resulting in the persistence of the single-ion Kondo effect down to low temperatures. Analogously, the fact that the canting angle value is essentially the same for higher concentrations of Y implies that a certain fraction of this angle "does not require" multiple Tb ions or periodicity of Tb sub-lattice; in other words, nearest neighbour interaction effects (which is referred to as 'local' in this case) contribute to this parameter.

**Conclusion**

We have carried out a detailed temperature dependent neutron diffraction studies on the Haldane spin-chin system $Tb_{2-x}Y_xBaNiO_5$ with the aim of throwing light on the intriguing observation of critical canting angle (of magnetic moments) for the onset of multiferroicity in $Tb_2BaNiO_5$. This angle in the multiferroic state is found to be significantly lowered for a small doping of Y for Tb, and this finding implies the contribution of a cooperative effect (among magnetic ions) to this characteristic angle in this system. The fact that the values are essentially independent of dopant concentration for higher doping implies a local contribution. This finding reveals that the factors deciding the critical canting angle has its origin in cooperative effects (between magnetic moments, meaning some kind of long-range) as well as in local (short-range) interactions. We hope that this conceptually novel proposal will help advance the theories in the field of multiferroics and also in engineering suitable multiferroic materials for applications.

**Acknowledgment:**
We acknowledge the support of Department of Atomic Energy, Government of India through Raja Ramanna Fellowship to E.V.S. This work was supported by Department of Atomic Energy, Government of India. (Grant no. RTI4003, DAE OM no. 1303/2/2019/R&D-II/ DAE/2079 dated 11.02.2020).

**References:**

[1] T. Kimura, T. Goto, H. Shintani, K. Ishiazaka, T. Arima, and Y. Tokura, Nature (London) 426, 55 (2003).
[2] G. Lawes, A. B. Harris, T. Kimura, N. Rogado, R. J. Cava, A. Aharony, O. Entin-Wohlman, T. Yildrim, M. Kenzelmann, C. Broholm, and A. P. Ramirez, Phys. Rev. Lett. 95, 087205 (2005).
[3] N. Hur, S. Park, P. A. Sharma, J. S. Ahn, S. Guha, and S.-W. Cheong, Nature (London) 429, 392 (2004).
[4] M. Fiebig, J. Phys. D: Appl. Phys. 38, R123 (2005).
[5] G. A. Smolenskii and I. Chupis, Sov. Phys. Usp. 25, 475 (1982).
[6] O. Heyer, N. Hollmann, I. Klassen, S. Jodlauk, L. Bohaty, P. Becker, J. A. Mydosh, T. Lorenz, and D. Khomskii, J. Phys.: Condens. Matter 18, L471 (2006).
[7] N. A. Spaldin, S. W. Cheong, and R. Ramesh, Phys. Today 63, 38 (2010).
[8] S. Dong, J.-M. Liu, S.-W. Cheong, and Z. Ren, Adv. Phys. 64, 519 (2015).
[9] M. Fiebig, T. Lottermoser, D. Meier, M. Trassin, Nat. Rev. Mater. 1, 16046 (2016).
[10] Choi Y, Okamoto J, Huang D, Chao K, Lin H, Chen C, Van Veenendaal M, Kaplan T and Cheong S, Phys. Rev. Lett. 102, 067601 (2009).
[11] Ram Kumar, Sanjay K Upadhyay, Y Xiao, W Ji and D Pal, J. Phys. D: Appl. Phys. 51, 385001 (2018).
[12] D. Khomskii, Physics 2, 20 (2009); J. van den Brink and D. I. Khomskii, J. Phys.: Condens. Matter 20, 434217 (2008).
[13] Y. J. Choi, H. T. Yi, S. Lee, Q. Huang, V. Kiryukhin, and S.-W. Cheong, Phys. Rev. Lett. 100, 047601 (2008).
[14] H. Katsura, N. Nagaosa, and A. V. Balatsky, Phys. Rev. Lett. 95, 057205 (2005).
[15] I. A. Sergienko and E. Dagotto, Phys. Rev. B 73, 094434 (2006).
[16] J. Hu, Phys. Rev. Lett. 100, 077202 (2008).

[17] A. B. Harris, T. Yildirim, A. Aharony, and O. Entin-Wohlman, Phys. Rev. B 73, 184433 (2006).
[18] T. Arima, J. Phys. Soc. Jpn. 76, 073702 (2007).
[19] T. A. Kaplan and S. D. Mahanti, Phys. Rev. B 83, 174432 (2011).
[20] F.D.M. Haldane, Continuum dynamics of the 1-D Heisenberg antiferromagnet: Identification with the O(3) nonlinear sigma model, Phys. Lett. 93A, 464 (1983).
[21] J. Darriet and L. P. Regnault, Solid State Commun. 86, 409 (1993).
[22] (a) K. Singh, T. Basu, S. Chowki, N. Mohapatra, K.K. Iyer, P.L. Paulose, E.V. Sampathkumaran, Phys Rev. B 88, 094438 (2013); (b) T. Basu, P.L. Paulose, K.K. Iyer, K. Singh, N. Mohapatra, S. Chowki, B. Gonde, E.V. Sampathkumaran, J. Phys. Condens. Matter. 26, 172202 (2014); (c) Tathamay Basu, K. Singh, N. Mohapatra, E.V. Sampathkumaran, J. Appl. Phys. 116, 114106 (2014) ;(d) S. Chowki, Tathamay Basu, K. Singh, N. Mohapatra, E.V. Sampathkumaran, J. Appl. Phys. 115 , 214107 (2014); (e) Tathamay Basu, V.V. Ravi Kishore, Smita Gohil, Kiran Singh, N. Mohapatra, S. Bhattacharjee, Babu Gonde, N.P. Lalla, Priya Mahadevan, Shankar Ghosh, E.V. Sampathkumaran, Sci. Rep. 4 , 5636 (2014); (f) Tathamay Basu, Niharika Mohapatra, Kartik K. Iyer, Kiran Singh, E.V. Sampathkumaran, AIP Adv. 5, 037128 (2015); (g) Sanjay Kumar Upadhyay, Kartik K. Iyer, E.V. Sampathkumaran, Phys. B Condens. Matter 524, 123–126 (2017); (h) Sanjay Kumar Upadhyay, E.V. Sampathkumaran, AIP Conf. Proc. 1942, 130061 (2018); (i) Sanjay K. Upadhyay, E.V. Sampathkumaran, S. Rayaprol, A. Hoser, Mater. Res. Express 6, 036107 (2019).
[23] (a) S.K. Upadhyay, P.L. Paulose, E.V. Sampathkumaran, Phys. Rev. B 96, 014418 (2017); (b) K.K. Iyer, Ram Kumar, S. Rayaprol, K. Maiti and E.V. Sampathkumaran, Phys. Rev. Mater. 5**,** 084401 (2021).
[24] Ram Kumar, Sudhindra Rayaprol, Sarita Rajput, Tulika Maitra, D.T. Adroja, Kartik K. Iyer, Sanjay K. Upadhyay, E.V. Sampathkumaran, Phys. Rev. B 99, 100406 (2019).

[25] Sanjay K. Upadhyay, E.V. Sampathkumaran, J. Phys. Condens. Matter 31, 39LT01 (2019).
[26] Sanjay K. Upadhyay, E.V. Sampathkumaran, J. Appl. Phys. 125, 174106 (2019).
[27] Ram Kumar, S. Rajput, T. Maitra, A. Hoser, S. Rayaprol, S. K. Upadhyay, K. K. Iyer, K. Maiti, and E. V. Sampathkumaran, J. Alloys Compd. 862, 158514 (2021);
[28] (a) R.E. Lechner, R.V. Wallpach, H.A. Graf, F.A. Kasper, L. Mokrani, Nuclear Instruments and Methods in Physics Research Section A: Accelerators, Spectrometers, Detectors and Associated Equipment 338, 65–70 (1994); (b) N. Stüßer, M. Hofmann, Nuclear Instruments and Methods in Physics Research Section A: Accelerators, Spectrometers, Detectors and Associated Equipment 482, 744–751 (2002); (c) A. Buchsteiner, N. Stüßer, Nuclear Instruments and Methods in Physics Research Section A: Accelerators, Spectrometers, Detectors and Associated Equipment 598, 534–541(2009).
[29] J. Rodriguez Carvajal, Physica B Condens. Matter 192, 55–69 (1993).
[30]. N.B. Brandt and V.V. Moshchalkov, Adv. Phys. 33, 373-468 (1984).

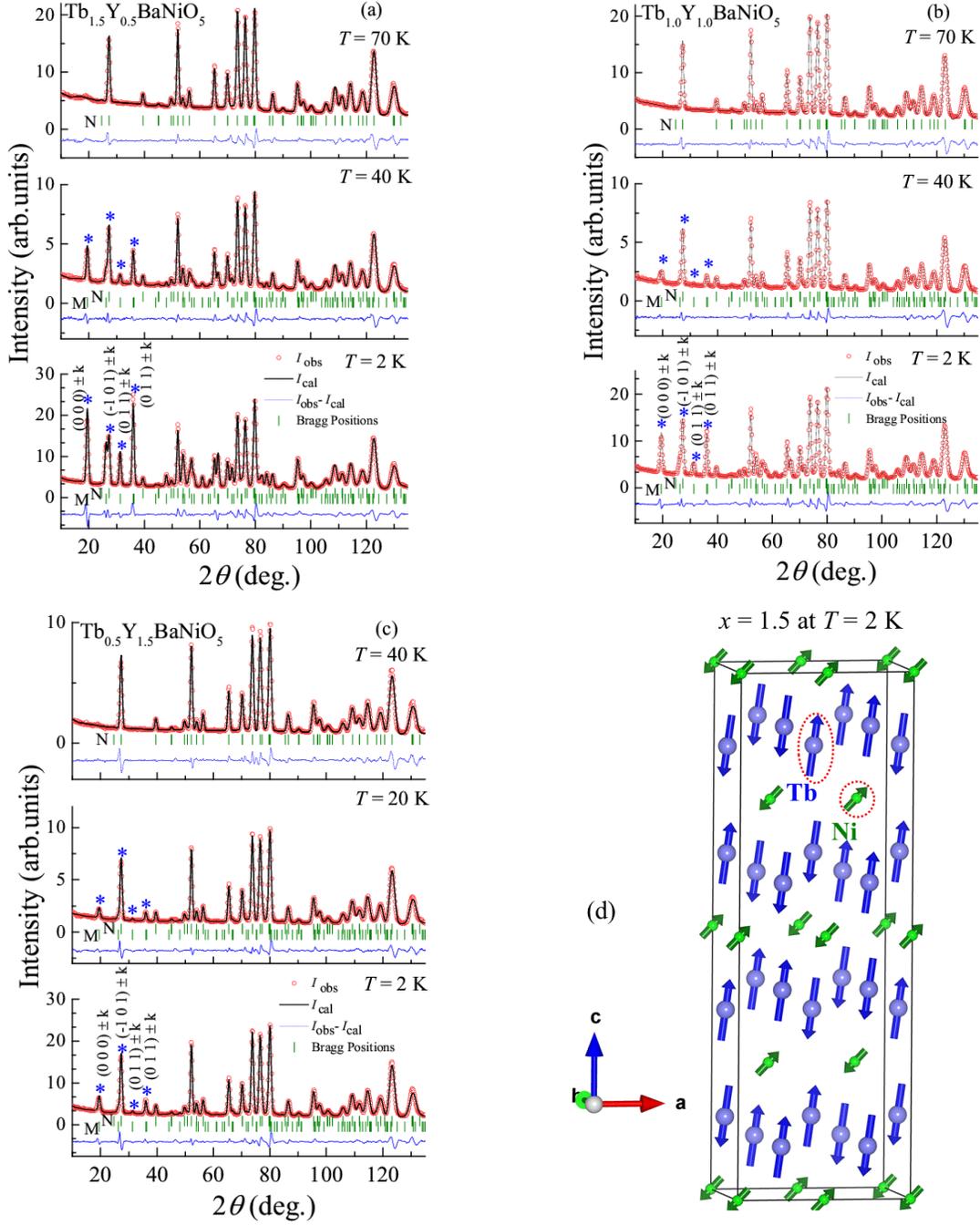

FIG. 1. (a-c) Rietveld refined neutron-diffraction patterns of Y substituted $Tb_2BaNiO_5$ at selected temperatures. For all three bottom panels, the (*hkl*) values for magnetic Bragg peaks with $k$ = (½, 0, ½) are shown (marked by asterisks). Nuclear (N) and magnetic (M) peak positions are shown by vertical green ticks. (d) depicts the magnetic structure of $Tb_{0.5}Y_{1.5}BaNiO_5$ at 2 K.

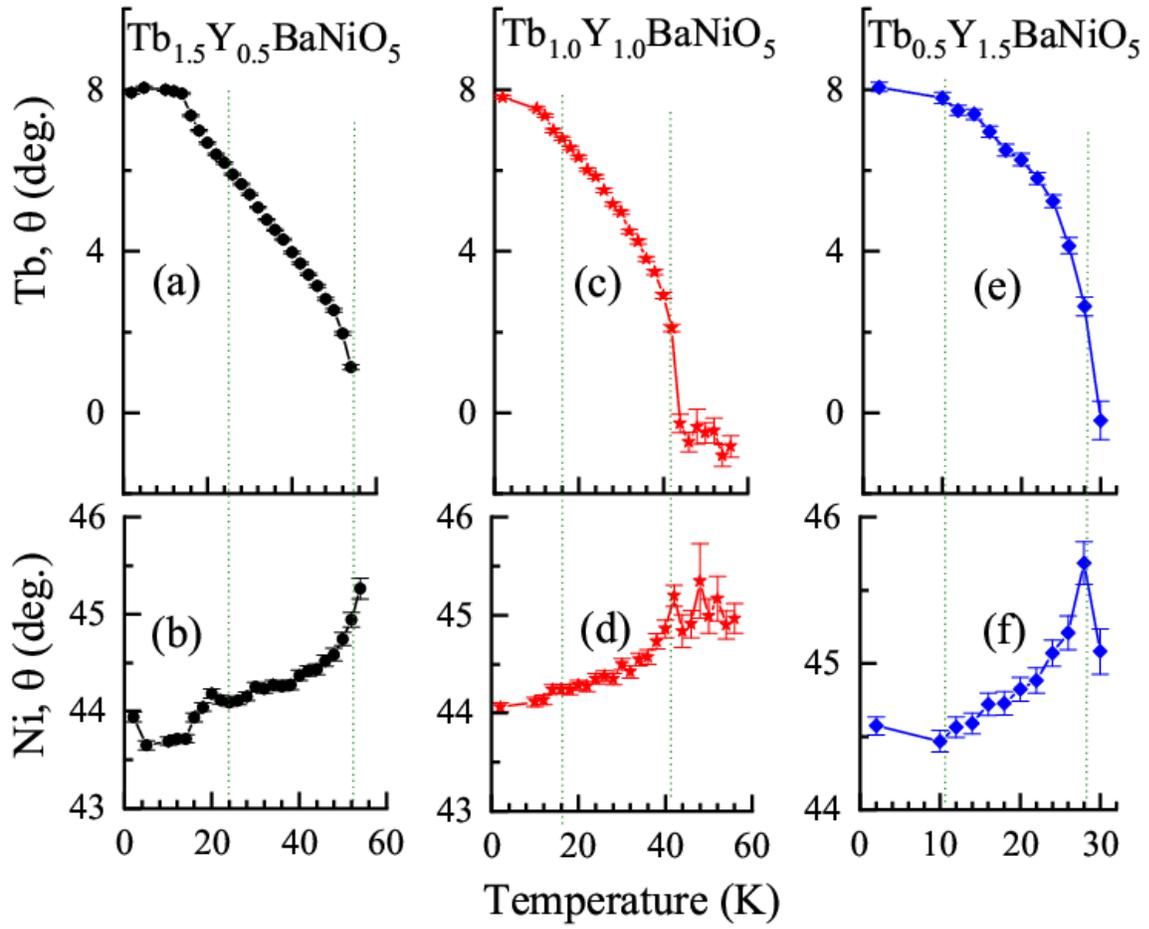

FIG. 2. Temperature dependence of canting angle with respect to c-axis for the magnetic moments of Tb (see *a*, *c*, and *e*) and Ni (see *b*, *d*, and *f*) for Y substituted $Tb_2BaNiO_5$. Vertical dotted lines are drawn at respective $T_{N1}$ and $T_{N2}$. A small moment a few degrees beyond $T_{N1}$ suggests persistence of short-range magnetic correlations.

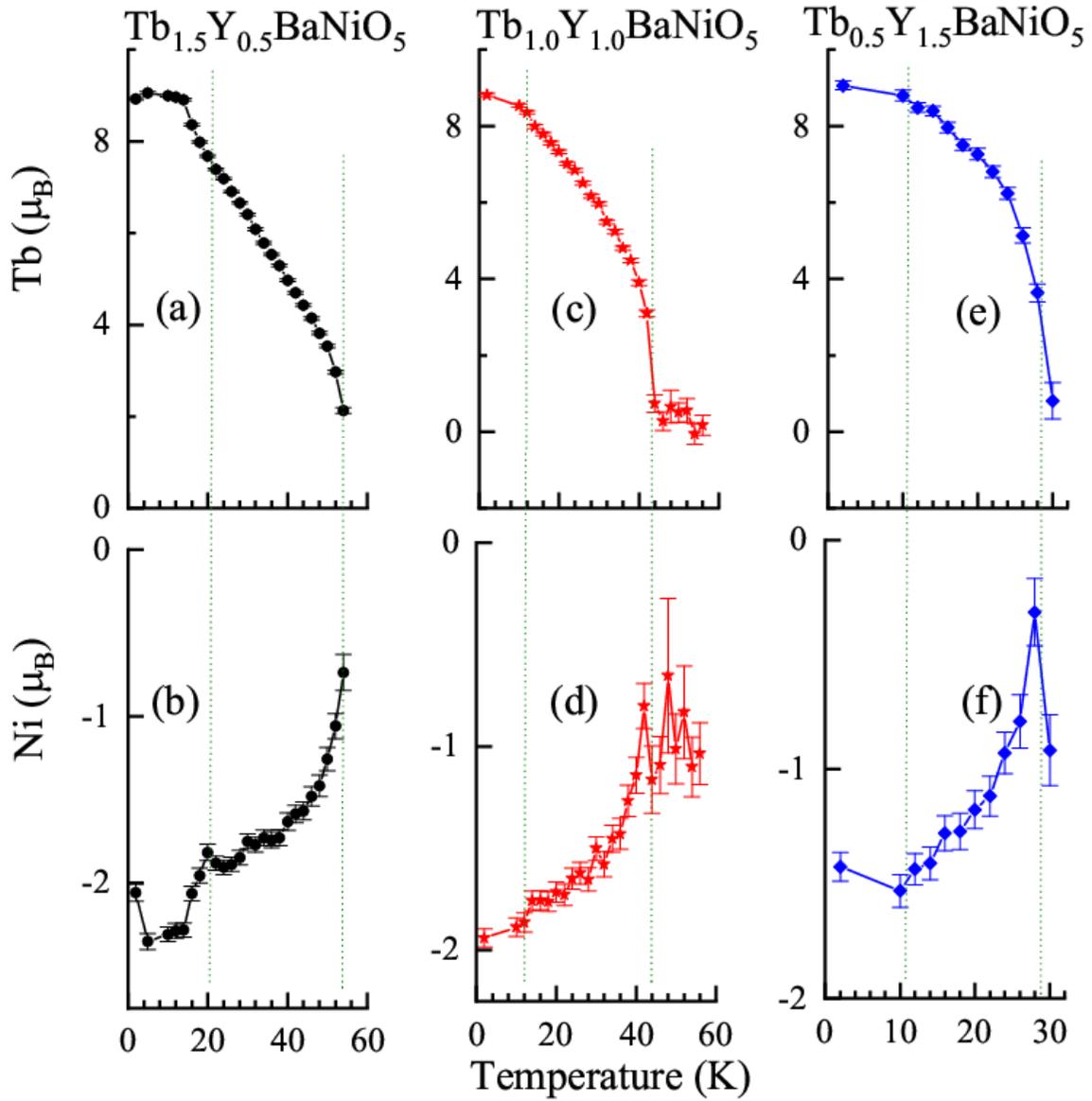

FIG. 3a Temperature dependence of magnetic moments of Tb "(see *a, c,* and *e)* and Ni (see *b, d,* and *f)* in Y substituted $Tb_2BaNiO_5$. Vertical dotted lines are drawn at respective $T_{N1}$ and $T_{N2}$.

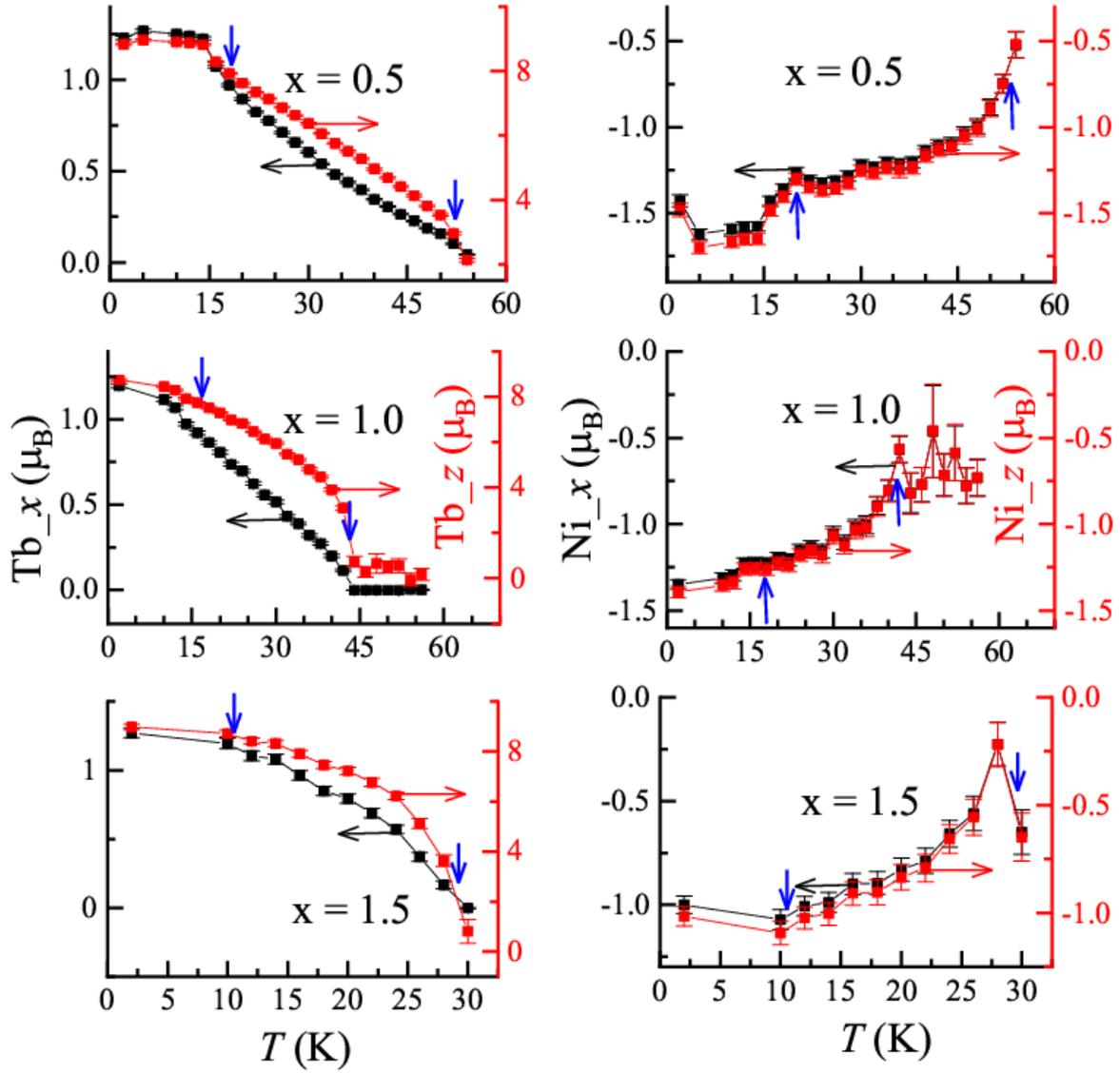

FIG. 3b: The *x* and *z* components of Tb and Ni magnetic moments as a function of temperature for $Tb_{2-x}Y_xBaNiO_5$. The vertical blue arrows mark respective characteristic magnetic transition temperatures discussed in the text.

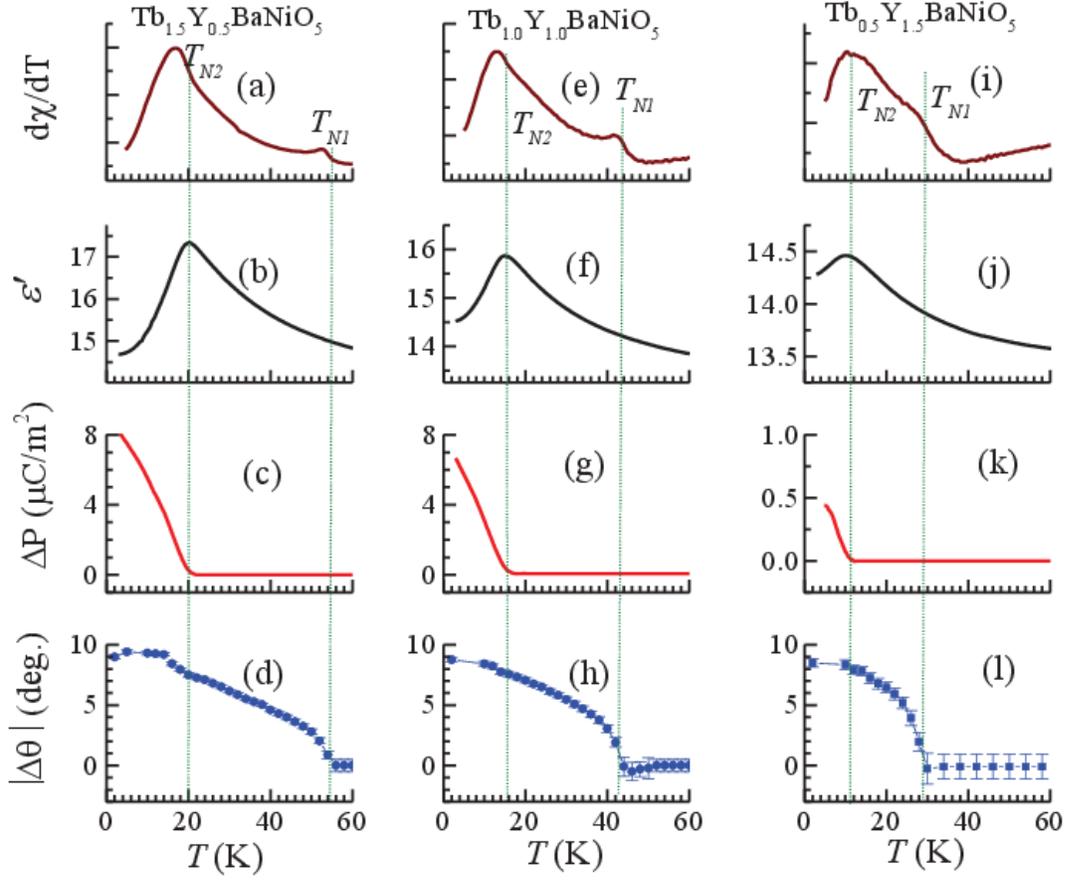

FIG. 4. Temperature dependence of magnetic susceptibility (as d$\chi$/dT), dielectric constants, change in electric polarization and relative canting angles of Tb and Ni magnetic moments in Y substituted Tb$_2$BaNiO$_5$ is shown in a-d, e-h, and i-l respectively. Vertical dotted lines are drawn at respective $T_{N1}$ and $T_{N2}$.